\documentclass[sigconf,screen]{acmart}
\makeatletter
\renewcommand\@formatdoi[1]{\ignorespaces}
\makeatother

\usepackage{booktabs}       % For formal tables
\usepackage{amsmath}        % math symbols
\usepackage{amssymb}        % more math symbols
\usepackage{amsthm}         % theorem-stuff
\usepackage{adjustbox}      % make floated boxes
\usepackage{xcolor}         % colors
\usepackage{textcomp}
\usepackage{braket}			% physics bra-ket notation (\Set{})
\usepackage{svg}            % support for vector graphics
\usepackage{url}            % support for \url{}
\usepackage{hyperref}       % hyperlinks

\newcommand{\sub}{\subseteq}                                    % better subset symbol
\newcommand{\bee}{\begin{equation}\begin{aligned}}              % display math with aligned eqns
\newcommand{\eee}{\end{aligned}\end{equation}}                  % end display
\newcommand{\fracc}{\frac}                                      % autocomplete frac
\renewcommand{\phi}{\varphi}                                    % prettier phi

\makeatletter
    {\@parboxrestore%
     \begin{adjustwidth}{}{\leftmargin}%
    }{\end{adjustwidth}
     }
\makeatother

\newif\iffinal
\iffinal
    \newcommand\kyle[1]{}
    \newcommand\luann[1]{}
    \newcommand\brendan[1]{}

\else
    \newcommand\kyle[1]{{\color{purple}[Kyle: #1]}}
    \newcommand\luann[1]{{\color{green}[Luann: #1]}}
    \newcommand\brendan[1]{{\color{blue}[Brendan: #1]}}

\fi

\setcopyright{rightsretained}
\acmDOI{}

\acmISBN{N/A}

\acmConference[SC'18]{ACM Student Research Competition}{November 2018}{Dallas, Texas USA}
\acmYear{2018}
\copyrightyear{2018}

\makeatletter
\def\@copyrightspace{\relax}
\makeatother

\hypersetup{
    colorlinks,             % make links pretty
    citecolor={red},        % better than neon green
} 

\begin{document}
\title{Measuring Swampiness: Quantifying Chaos in Large Heterogeneous Data Repositories}
\subtitle{Extended Abstract}

\author{Luann Jung}
\affiliation{%
  \institution{Massachusetts Institute of Technology}
  \streetaddress{}
  \city{}
  \state{}
  \postcode{}
}
\email{luju@mit.edu}

\author{Brendan Whitaker}
\affiliation{%
  \institution{Ohio State University}
  \streetaddress{}
  \city{}
  \state{}
  \postcode{}
}
\email{whitaker.213@osu.edu}

\author{Kyle Chard (advisor)}
\affiliation{%
  \institution{University of Chicago}
  \streetaddress{}
  \city{}
  \country{}}
\email{chard@uchicago.edu}

\author{Aaron J. Elmore (advisor)}
\affiliation{%
  \institution{University of Chicago}
  \city{}
  \country{}
}
\email{aelmore@cs.uchicago.edu}

\renewcommand{\shortauthors}{Measuring Swampiness: Quantifying Chaos in Large Heterogeneous Data Repositories}

\begin{abstract}
As scientific data repositories and filesystems grow in size and complexity, they become increasingly disorganized. The coupling of massive quantities of data with poor organization makes it challenging for scientists to locate and utilize relevant data, thus slowing the process of analyzing data of interest. To address these issues, we explore an automated clustering approach for quantifying the organization of data repositories. Our parallel pipeline processes heterogeneous filetypes (e.g., text and tabular data), automatically clusters files based on content and metadata similarities, and computes a novel ``cleanliness'' score from the resulting clustering. We demonstrate the generation and accuracy of our cleanliness measure using both synthetic and real datasets, and conclude that it is more consistent than other potential cleanliness measures.
\end{abstract}

\maketitle

\section{Introduction}

Traditional modes of organizing data repositories and filesystems are increasingly ineffective due to the size, heterogeneity, and complexity of data. Researchers are now turning to alternative organizational models such as data lakes---repositories for large quantities of raw data that are integrated in a pay-as-you-go fashion~\cite{jeffery2008pay, madhavan2007web}. However, users are often unwilling to spend time describing and organizing data, causing repositories to become opaque ``data swamps''\cite{hai2016constance} with poor metadata and confusing directory structures. 

To combat this problem, we propose a set of tools that automate the process of identifying content-based relationships between files. We present a parallel pipeline that crawls repositories, collects key information regarding data composition and distribution, and automatically clusters files based on extracted content and metadata. Our unsupervised clustering models aim to detect latent similarities in file subject, provenance, or purpose \cite{brackenbury2018} and then clusters accordingly. We use these clusters to define a novel ``cleanliness'' measure to quantify the organization of the data repository. This measure consists of a newly proposed frequency drop score which takes into account the directory composition and density of clusters generated by the pipeline. We explore the efficacy of our approach using synthetic data as well as a real-world climate science dataset~\cite{CDIAC}. 

\section{Methodology}

We implement a clustering-based pipeline to identify similar data irrespective of how it is organized. The pipeline is composed of four major steps:  crawling, preprocessing, clustering, and calculating cleanliness.

\begin{center}
\begin{figure}[!htbp]
\includegraphics[width=0.4\textwidth]{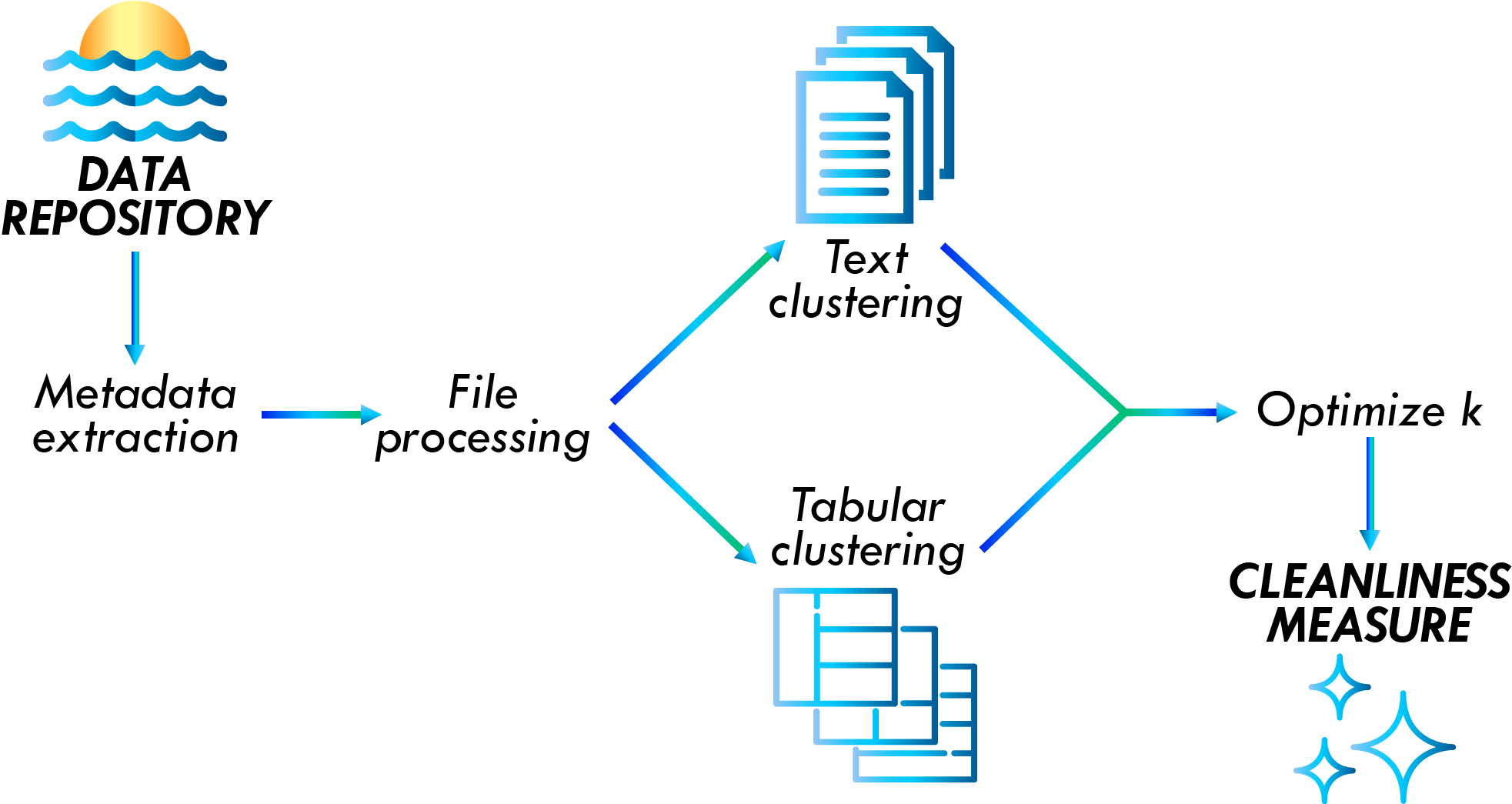}
\setlength{\belowcaptionskip}   {-18pt}
\caption{\label{pipeline} Clustering pipeline.}
\end{figure}
\end{center}

We focus on two data types: unstructured text and structured tabular data. 
First, we convert files into common formats (\texttt{.txt}/\texttt{.csv}). Then, we preprocess file contents according to their data type. Text data are tokenized, stemmed, and vectorized into a TF-IDF matrix, while schemas are extracted from tabular data and used to compute a pairwise Jaccard distance matrix.

For text files, we implement classic $k$-means clustering and the faster MiniBatch $k$-means clustering. For tabular files, we use agglomerative hierarchical clustering since
it does not rely on centroids or other features of Euclidean space. After clustering both filetypes, we generate output clusters, composition statistics, and a dataset cleanliness score. The pipeline is then repeated over a user-specified range of $k$ values to optimize the $k$ which best represents the data.

To measure cleanliness, we first define the frequency drop score for a clustering of some dataset $A$ by examining the distribution of directories constituting each cluster $C_i$. Given the number of files from each directory in a cluster, we identify the location of the largest ``frequency drop''---representing the point where the tail of the distribution begins. Let $\Set{D_1,...,D_m}$ be the set of all directories containing files from cluster $C_i \sub A$. We define the head $H_i$ as the set of all directories before the drop, and the tail $T_i$ as the set of all remaining directories of $C_i$. Under the assumptions that similar data are physically close in well-organized datasets and that the clustering $C = \Set{C_1,...,C_k}$ is sufficiently cohesive, the function $\mathcal{S}(C)$ yields a value in $[0,1]$ representing the cleanliness of the dataset. 
We define a logarithm-like function which is well-defined for a base of $1$:
\bee
\sigma(a,b) = 
\begin{cases}
\log_ab & \text{if }a > 1\\
0 & \text{if }a = 1
\end{cases}.
\eee

\noindent
The frequency drop score for each cluster is given by
\bee
\mathrm{drop}(C_i) = 
\begin{cases}
\fracc{1 - \sigma(m - 1, |H_i|)}{|C_i|}\sum\limits_{D_j \in H_i}|D_j| & \text{ if }m > 1\\
1 & \text{ otherwise}
\end{cases}, 
\eee
and the score for the entire clustering is given by
\bee
\mathcal{S}(C) = \fracc{|C_i|}{|A|}\sum_{i = 1}^k \mathrm{drop}(C_i). 
\eee

\section{Evaluation}
We evaluate our approach using synthetic data as well as the Carbon Dioxide Information and Analysis Center's (\texttt{CDIAC}) data repository. 

As a baseline, we generated three synthetic datasets based on $N$-ary trees. Each synthetic dataset includes one parent directory (root node) with $N$ children, each of which has $N$ children, extended to any chosen height $h$. Each leaf node contains twenty \texttt{.txt} files and twenty \texttt{.csv} files, with each file containing the same word repeated 100 times. Each word is unique to its leaf node, such that the number of expected clusters is equal to the number of leaf nodes. These datasets, when run through our pipeline, yield:
\begin{itemize}
    \item perfect clusters where each cluster contains only and all of the files with the same word.
    \item a cleanliness score of 1.0.
\end{itemize} 
With this as a baseline, we then shuffled the datasets such that files were randomly assigned to leaf directories. Table~\ref{synthetictable} shows that the cleanliness scores decrease as the dataset is shuffled.

\begin {table}[!htbp]
\begin{center}
%\begin{tabular}{c|cccccc}      
%\begin{tabular}{lllllll}       % left aligns, no bar
\begin{tabular}{@{}lllllll@{}}  % top/mid/bottom rule flush to text
\toprule
  & \multicolumn{6}{c}{\% Scrambled} \\
 Dataset & 0\% & 20\% & 40\% & 60\% & 80\% & 100\%\\
\midrule
 2-ary, \hspace{0.5mm} 5-height & 1.000 & 0.806 & 0.619 & 0.420 & 0.227 & 0.093\\ 
 3-ary, \hspace{0.5mm} 3-height & 0.963 & 0.765 & 0.595 & 0.429 & 0.188 & 0.079\\ 
 6-ary, \hspace{0.5mm} 2-height & 1.000 & 0.792 & 0.593 & 0.451 & 0.225 & 0.106\\ 
 40-ary, 1-height & 0.950 & 0.780 & 0.579 & 0.341 & 0.217 & 0.109\\ 
 \bottomrule
\end{tabular}
\caption{\label{synthetictable} Cleanliness scores for shuffled synthetic datasets.}
\end{center}
\vspace{-1.0cm}
\end{table}

\begin{center}
\begin{figure}[!htbp]
\includegraphics[width=0.47\textwidth]{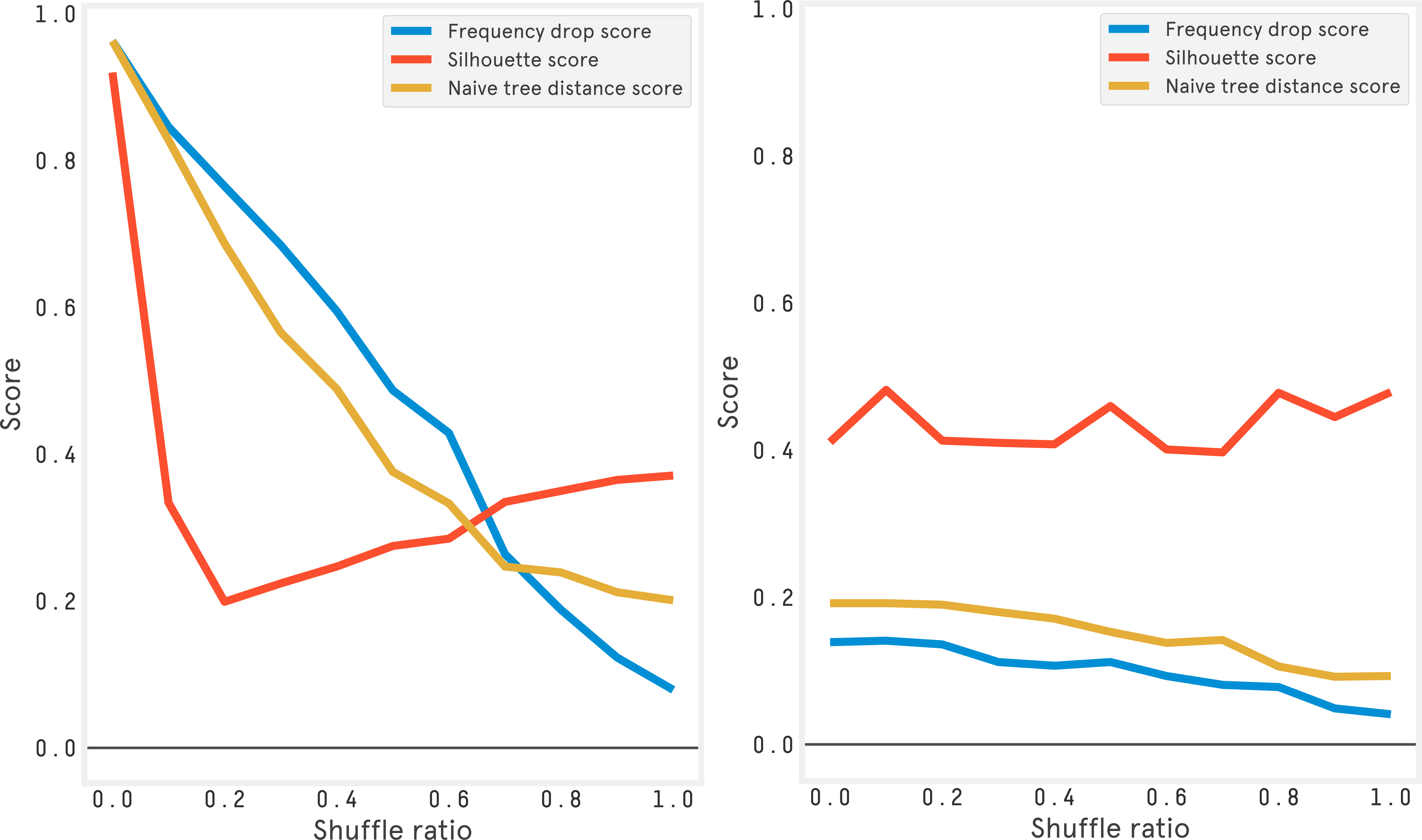}
\setlength{\belowcaptionskip}{-20pt}
\caption{\label{cleanlinessgraphs} Comparison of cleanliness measures - \textmd{$3$-ary tree synthetic dataset of tabular files with height $2$ (left), and tabular files from \texttt{pub8} (right). }}
\end{figure}
\end{center}

We compared our cleanliness score with two other measures: cluster cohesion and a modified Silhouette score~\cite{silhouette}, both computed with na{\"{\i}}ve filesystem tree distance. 
Figure \ref{cleanlinessgraphs} shows these measures calculated on progressively more shuffled synthetic datasets and real scientific data (from the \texttt{pub8} subset of \texttt{CDIAC}). We conclude that the silhouette scores are inconsistent and noisy when compared to our cleanliness measure. The na{\"{\i}}ve tree distance score is comparable, but still fails to discriminate between repositories with vastly different organizational structures in some adversarial examples.

\section{Summary}
We introduce a parallel pipeline for automated content-based clustering of files from large heterogeneous data repositories. These clusters are then used to derive a novel measure of the organizational cleanliness of a repository. The measure we developed exhibits better consistency than existing measures when tested on a variety of datasets. The code for our pipeline is available here: 
\url{https://github.com/lollyluann/cluster-datalake}

\nocite{*}  % if this is removed, bibtex breaks. no idea why
\bibliographystyle{ACM-Reference-Format}
\bibliography{final-bibliography}


\begin{thebibliography}{8}

%%% the Web address.

\ifx \showCODEN    \undefined \def \showCODEN     #1{\unskip}     \fi
\ifx \showDOI      \undefined \def \showDOI       #1{#1}\fi
\ifx \showISBNx    \undefined \def \showISBNx     #1{\unskip}     \fi
\ifx \showISBNxiii \undefined \def \showISBNxiii  #1{\unskip}     \fi
\ifx \showISSN     \undefined \def \showISSN      #1{\unskip}     \fi
\ifx \showLCCN     \undefined \def \showLCCN      #1{\unskip}     \fi
\ifx \shownote     \undefined \def \shownote      #1{#1}          \fi
\ifx \showarticletitle \undefined \def \showarticletitle #1{#1}   \fi
\ifx \showURL      \undefined \def \showURL       {\relax}        \fi
% invisible to TeX
\providecommand\bibfield[2]{#2}
\providecommand\bibinfo[2]{#2}
\providecommand\natexlab[1]{#1}
\providecommand\showeprint[2][]{arXiv:#2}

\bibitem[\protect\citeauthoryear{Beckman, Skluzacek, Chard, and Foster}{Beckman
  et~al\mbox{.}}{2017}]%
        {beckman2017skluma}
\bibfield{author}{\bibinfo{person}{Paul Beckman}, \bibinfo{person}{Tyler~J
  Skluzacek}, \bibinfo{person}{Kyle Chard}, {and} \bibinfo{person}{Ian
  Foster}.} \bibinfo{year}{2017}\natexlab{}.
\newblock \showarticletitle{Skluma: A statistical learning pipeline for taming
  unkempt data repositories}. In \bibinfo{booktitle}{\emph{29th International
  Conference on Scientific and Statistical Database Management}}.
  \bibinfo{pages}{41}.
\newblock


\bibitem[\protect\citeauthoryear{Brackenbury, Liu, Mondal, Elmore, Ur, Chard,
  and Franklin}{Brackenbury et~al\mbox{.}}{2018}]%
        {brackenbury2018}
\bibfield{author}{\bibinfo{person}{Will Brackenbury}, \bibinfo{person}{Rui
  Liu}, \bibinfo{person}{Mainack Mondal}, \bibinfo{person}{Aaron~J. Elmore},
  \bibinfo{person}{Blase Ur}, \bibinfo{person}{Kyle Chard}, {and}
  \bibinfo{person}{Michael~J. Franklin}.} \bibinfo{year}{2018}\natexlab{}.
\newblock \showarticletitle{Draining the Data Swamp: A Similarity-based
  Approach}. In \bibinfo{booktitle}{\emph{Proceedings of the Workshop on
  Human-In-the-Loop Data Analytics}} \emph{(\bibinfo{series}{HILDA'18})}.
  \bibinfo{publisher}{ACM}, \bibinfo{address}{New York, NY, USA}, Article
  \bibinfo{articleno}{13}, \bibinfo{numpages}{7}~pages.
\newblock
\showISBNx{978-1-4503-5827-9}
\urldef\tempurl%
\url{https://doi.org/10.1145/3209900.3209911}
\showDOI{\tempurl}


\bibitem[\protect\citeauthoryear{Chessell, Scheepers, Nguyen, van Kessel, and
  van~der Starre}{Chessell et~al\mbox{.}}{2014}]%
        {chessell:bigdata}
\bibfield{author}{\bibinfo{person}{M. Chessell}, \bibinfo{person}{F.
  Scheepers}, \bibinfo{person}{N. Nguyen}, \bibinfo{person}{R. van Kessel},
  {and} \bibinfo{person}{R. van~der Starre}.} \bibinfo{year}{2014}\natexlab{}.
\newblock \bibinfo{booktitle}{\emph{Governing and Managing Big Data for
  Analytics and Decision Makers}}.
\newblock
\urldef\tempurl%
\url{http://www.redbooks.ibm.com/redpapers/pdfs/redp5120.pdf}
\showURL{%
\tempurl}


\bibitem[\protect\citeauthoryear{Hai, Geisler, and Quix}{Hai
  et~al\mbox{.}}{2016}]%
        {hai2016constance}
\bibfield{author}{\bibinfo{person}{Rihan Hai}, \bibinfo{person}{Sandra
  Geisler}, {and} \bibinfo{person}{Christoph Quix}.}
  \bibinfo{year}{2016}\natexlab{}.
\newblock \showarticletitle{Constance: An intelligent data lake system}. In
  \bibinfo{booktitle}{\emph{Proceedings of the 2016 International Conference on
  Management of Data}}. ACM, \bibinfo{pages}{2097--2100}.
\newblock


\bibitem[\protect\citeauthoryear{Jeffery, Franklin, and Halevy}{Jeffery
  et~al\mbox{.}}{2008}]%
        {jeffery2008pay}
\bibfield{author}{\bibinfo{person}{Shawn~R Jeffery}, \bibinfo{person}{Michael~J
  Franklin}, {and} \bibinfo{person}{Alon~Y Halevy}.}
  \bibinfo{year}{2008}\natexlab{}.
\newblock \showarticletitle{Pay-as-you-go user feedback for dataspace systems}.
  In \bibinfo{booktitle}{\emph{Proceedings of the 2008 ACM SIGMOD international
  conference on Management of data}}. ACM, \bibinfo{pages}{847--860}.
\newblock


\bibitem[\protect\citeauthoryear{Madhavan, Jeffery, Cohen, Dong, Ko, Yu, and
  Halevy}{Madhavan et~al\mbox{.}}{2007}]%
        {madhavan2007web}
\bibfield{author}{\bibinfo{person}{Jayant Madhavan}, \bibinfo{person}{Shawn~R
  Jeffery}, \bibinfo{person}{Shirley Cohen}, \bibinfo{person}{Xin Dong},
  \bibinfo{person}{David Ko}, \bibinfo{person}{Cong Yu}, {and}
  \bibinfo{person}{Alon Halevy}.} \bibinfo{year}{2007}\natexlab{}.
\newblock \showarticletitle{Web-scale data integration: You can only afford to
  pay as you go}. CIDR.
\newblock


\bibitem[\protect\citeauthoryear{Rousseeuw}{Rousseeuw}{1987}]%
        {silhouette}
\bibfield{author}{\bibinfo{person}{Peter~J. Rousseeuw}.}
  \bibinfo{year}{1987}\natexlab{}.
\newblock \showarticletitle{Silhouettes: A graphical aid to the interpretation
  and validation of cluster analysis}.
\newblock \bibinfo{journal}{\emph{J. Comput. Appl. Math.}}
  \bibinfo{volume}{20} (\bibinfo{year}{1987}), \bibinfo{pages}{53 -- 65}.
\newblock
\showISSN{0377-0427}
\urldef\tempurl%
\url{https://doi.org/10.1016/0377-0427(87)90125-7}
\showDOI{\tempurl}


\bibitem[\protect\citeauthoryear{{U.S. Dept. of Energy}}{{U.S. Dept. of
  Energy}}{2017}]%
        {CDIAC}
\bibfield{author}{\bibinfo{person}{{U.S. Dept. of Energy}}.}
  \bibinfo{year}{2017}\natexlab{}.
\newblock \bibinfo{title}{{Carbon Dioxide Information Analysis Center}}.
\newblock
\newblock
\newblock
\shownote{\url{ftp://cdiac.ornl.gov}. Visited Feb. 28, 2017.}


\end{thebibliography}

\end{document}